\documentclass {amsart}
\usepackage[pdftex]{graphicx}

\newcommand{\degF}{ ^\circ \mbox{F} }

\renewcommand{\xi}{{a}}
\renewcommand{\eta}{{b}}

\title{Local Warming}
\author{ Robert J. Vanderbei}
\thanks{Department of Operations Research and Financial Engineering,
	Princeton University,
	Princeton, NJ 08544 ({\tt rvdb@princeton.edu}).
}

\begin{document} 

\maketitle
 
\begin{abstract}  
Using 55 years of daily average temperatures from a local weather station, I
made a least-absolute-deviations (LAD) regression model that accounts for three
effects: seasonal variations, the 11-year solar cycle, and a linear trend.
The model was formulated as a linear programming problem and solved using widely 
available optimization software.
The solution indicates that temperatures have gone up by about $2\degF$ over the 55
years covered by the data.  
It also correctly identifies the known phase of the solar cycle; 
i.e., the date of the last solar minimum.
It turns out that the maximum slope of the 
solar cycle sinusoid in the regression model is about the 
same size as the slope produced by the linear trend.
The fact that the solar cycle was correctly extracted by the model is a strong indicator
that effects of this size, in particular the slope of the linear trend, can be accurately
determined from the 55 years of data analyzed.

The main purpose for doing this analysis is to demonstrate that it is easy to find and 
analyze archived temperature data for oneself.  In particular, this problem makes a good class
project for upper-level undergraduate courses in optimization or in statistics.

It is worth noting that a similar least-squares model failed to
characterize the solar cycle correctly and hence even though it too indicates that temperatures 
have been rising locally, one can be less confident in this result.

The paper ends with a section presenting similar results from a few thousand sites distributed
world-wide, some results from a modification of the model that includes both
temperature and humidity,
as well as a number of suggestions for future work and/or ideas for enhancements that could be
used as classroom projects.

\end{abstract}



\section{Introduction}

Most research on climate change aims to produce a high-fidelity model of climate
that spans centuries \cite{MBH98} if not millennia \cite{JBBT98}.   Since
directly observed temperature data is not available over these time scales, such models 
are forced to resort to proxy climate indicators
(see, e.g., \cite{HSHSCJ96}).
In this paper, actual temperature readings from a single undisturbed location
spanning a time horizon of $55$ years are analyzed
using a {\em least-absolute deviations (LAD)} regression model that 
robustly extracts a small linear trend from the much larger seasonal
variations.
An analogous least-squares regression model generates results that are less
reliable than those obtained with the LAD model.

The purpose of this paper is not to attempt to improve on any of the global
warming estimates that exist in the literature.  
Rather,
the main purpose of this paper is to show that it is fairly easy to find
and analyze archived temperature data with the hope that many others will
make a similar analysis for places that are of interest to them.   I also hope
to inspire educators who teach courses in optimization and/or statistics to
develop classroom projects based on the ideas presented here.

Least absolute deviations regression \cite{BS83} belongs to a class of statistical
techniques called {\em robust statistics} \cite{HR09}.  The sample median is the
simplest and most widely used example of a robust statistic.   Sample medians
have played an important role in a wide range of scientific fields including
astrophysics \cite{GVPR01}, medicine \cite{CG88}, and signal processing
\cite{BHM83} to name a few.
A secondary purpose of this paper is to demonstrate that, at least for the data considered
here, least-absolute deviations regression provides better results than the
corresponding least-squares regression.

\section{The Data} 

The {\em National Oceanic and Atmospheric Administration} (NOAA) collects and
archives weather data from thousands of collection sites around the globe.  The
data format and instructions for downloading the data can be found on the
NOAA website \cite{noaa_readme}
%
%
as can a list of the roughly 9000 weather stations \cite{noaa_stationlist}.
%
%
McGuire Air Force Base, located not far from Princeton NJ, 
is one of the archived weather stations.   
This particular weather station seemed good for a number of reasons:  
(i) it is about 50 miles from New York City and 30 miles from Philadelphia,  
(ii) it is in a rather undeveloped part of the state,  
(iii) it was established in 1937 and has been a major airbase since 1942, and
finally,
(iv) it is only about 25 miles from the Atlantic Ocean and therefore its climate
should be moderated somewhat by its proximity to an ocean.

\begin{figure}
\begin{center}
\includegraphics[width=4.5in]{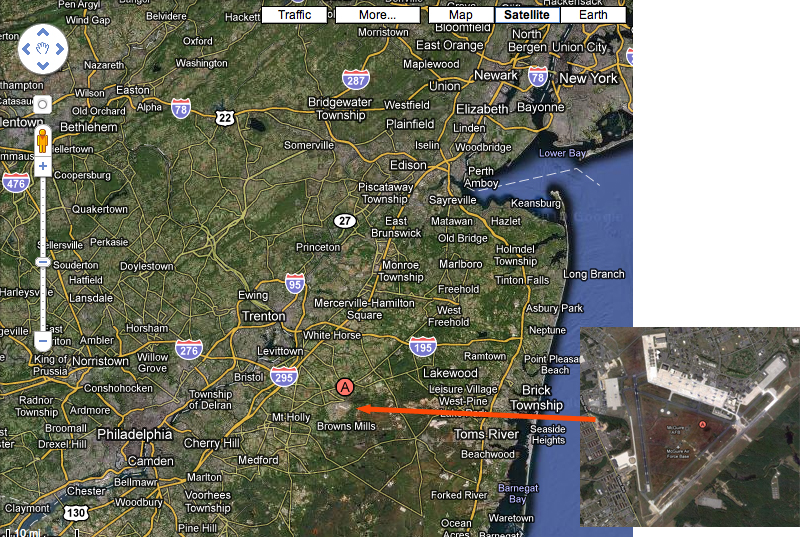}
\end{center}
\caption{McGuire Air Force Base}
\vspace*{0.3in}
\end{figure}

\section{The Model} 
Let $T_d$ denote the average temperature in degrees Fahrenheit on day $d \in D$
where $D$ is the set of days from January 1, 1955, to August 13, 2010 (that's 20,309 days).  

We wish to model the average temperature as a {\em constant} $x_0$ plus a {\em
linear trend} $x_1 d$
plus a sinusoidal function with a one-year period representing {\em seasonal
changes},
\[
	x_2 \cos( 2 \pi d/365.25) + x_3 \sin( 2 \pi d/365.25),
\]
plus a sinusoidal function with a period of $10.7$ years to represent the {\em solar cycle},
\[
	  x_4 \cos( 2 \pi d/(10.7 \times 365.25)) 
	+ x_5 \sin( 2 \pi d/(10.7 \times 365.25)).
\]
The parameters $x_0, x_1, \ldots, x_5$ are unknown regression coefficients. 
Our aim is to find the values of these parameters that
minimize the sum of the absolute deviations:
\begin{equation}
   \begin{array}{ll}
   \displaystyle
   \min_{x_0, \ldots, x_5} \sum_{d \in D} 
        & \left| x_0 + x_1 d \right. \\[-0.1in]
	& \quad + x_2 \cos( 2 \pi d/365.25) 
	        + x_3 \sin( 2 \pi d/365.25) \\[0.1in]
	& \quad + x_4 \cos( 2 \pi d/(10.7 \times 365.25)) 
	        + x_5 \sin( 2 \pi d/(10.7 \times 365.25))  \\[0.1in]
	& \quad \left. - T_d \right| .
   \end{array} \label{55}
\end{equation}

We use the usual trick of introducing a new variable for each absolute value
term and then adding a pair of constraints that say that this new variable
dominates the expression that was inside the absolute values and its negative.
The result is a {\em linear programming (LP)} problem.
One can check that the solution to the LP formulation is identical to the
solution to the original model whenever the original problem is ``convex''.  In
particular, they are the same whenever one minimizes a nonnegative weighted sum of absolute values 
of linear expressions subject to linear equality and inequality constraints
(\cite{Van07}, Chapter 12).

The linear programming problem, expressed in the {\sc ampl}
modeling language \cite{FGK93}, is shown in Figure \ref{fig2}.
{\sc ampl} models, with their associated user-supplied data sets, can be solved online
using the {\em Network Enabled Optimzation Server (NEOS)} at Argonne National
Labs \cite{NEOS}.

A {\em least-absolute-deviations} (LAD) model was chosen instead of a least-squares
model because LAD regression, like the median statistic, is insensitive to
``outliers'' in the data.  The least-squares variant is discussed in Section
\ref{sec-lsr}.

\begin{figure}
\scriptsize
\begin{center}
\begin{verbatim}
set DATES ordered;
param avg {DATES};
param day {DATES};
param pi := 4*atan(1);

var x {j in 0..5};
var dev {DATES} >= 0, := 1;

minimize sumdev: sum {d in DATES} dev[d];
subject to def_pos_dev {d in DATES}:
    x[0] + x[1]*day[d] + x[2]*cos( 2*pi*day[d]/365.25)
                       + x[3]*sin( 2*pi*day[d]/365.25)
                       + x[4]*cos( 2*pi*day[d]/(10.7*365.25))
                       + x[5]*sin( 2*pi*day[d]/(10.7*365.25))
        - avg[d] 
    <= dev[d];
subject to def_neg_dev {d in DATES}:
    -dev[d] <= 
    x[0] + x[1]*day[d] + x[2]*cos( 2*pi*day[d]/365.25)
                       + x[3]*sin( 2*pi*day[d]/365.25)
                       + x[4]*cos( 2*pi*day[d]/(10.7*365.25))
                       + x[5]*sin( 2*pi*day[d]/(10.7*365.25))
        - avg[d];

data;

set DATES := include "data/Dates.dat";
param: avg := include "data/McGuireAFB.dat";
let {d in DATES} day[d] := ord(d,DATES);

solve;

\end{verbatim}
\end{center}
\normalsize
\caption{The model expressed in the AMPL modeling language.}
\label{fig2}
\vspace*{0.3in}
\end{figure}

\subsection{Confidence Intervals} 

Let $x_0^*, \ldots, x_5^*$ denote the optimal solution to \eqref{55} and let
$\varepsilon_d$ denote the corresponding deviations:
\begin{eqnarray*}
    \varepsilon_d 
    & = &
   \displaystyle
          x_0^* + x_1^* d 
	+ x_2^* \cos( 2 \pi d/365.25) 
	        + x_3^* \sin( 2 \pi d/365.25) \\[0.1in]
	&& \quad + x_4^* \cos( 2 \pi d/(10.7 \times 365.25)) 
	        + x_5^* \sin( 2 \pi d/(10.7 \times 365.25))  
	- T_d .
\end{eqnarray*}
In the well-known {\em bootstrap} method \cite{Efr79}, we assume that these $\varepsilon_d$'s
form an empirical distribution associated with an unknown underlying error
distribution.   As such, we can sample from these errors and generate new
data:
\begin{eqnarray*}
    T_d'
    & = &
   \displaystyle
          x_0^* + x_1^* d 
	+ x_2^* \cos( 2 \pi d/365.25) 
	        + x_3^* \sin( 2 \pi d/365.25) \\[0.1in]
	&& \quad + x_4^* \cos( 2 \pi d/(10.7 \times 365.25)) 
	        + x_5^* \sin( 2 \pi d/(10.7 \times 365.25))  
		- \varepsilon_d' 
\end{eqnarray*}
where the $\varepsilon_d'$ are drawn independently and with replacement from the
set $\{ \varepsilon_d : d \in D \}$.  In this manner we can generate several
alternate data sets that are statistically similar to the original one and we can
then recompute the parameters $x_0, \ldots, x_5$ using these replicated data
sets to get multiple estimates for each parameter and thereby 
compute $2\sigma$-confidence intervals for each parameter (or any other derived
parameter).

\section{The Results} 
The linear programming problem can be solved 
in only a few minutes on a modern laptop computer.   
The optimal values of the parameters together with their $2\sigma$-confidence
intervals are

\begin{eqnarray*}
x_0 &=& \phantom{-}52.6\phantom{00} \pm 0.27 ~\degF \\
x_1 &=& \phantom{-0}3.63\phantom{0} \pm 0.75 ~^\circ \mbox{F/century}\\
x_2 &=&-20.4\phantom{00} \pm 0.16 ~\degF \\
x_3 &=& \;\;-8.31\phantom{0} \pm 0.17 ~\degF \\
x_4 &=& \;\;-0.197 \pm 0.137 ~\degF \\
x_5 &=& \phantom{-0}0.211 \pm 0.202 ~\degF \\
\end{eqnarray*}


\subsection{Linear Trend}

From $x_0$, we see that the nominal temperature at McGuire AFB 
was $52.56 \pm 0.27 ~\degF$ (on January 1, 1955).  

We also see, from $x_1$, that there is a positive trend of 
$0.000099 \pm 0.000020 ~^\circ \mbox{F/day}$, 
which scales to $3.63 \pm 0.75 ~\degF$ per
century.  This result is consistent with results from global climate change models,
which predict a per century warming rate of between 
$2.0 ~^\circ \mbox{C}$ and $2.4 ~^\circ \mbox{C}$ 
(\cite{KKB00,ipcc}).

\subsection{Amplitude of the Sinusoidal Fluctuations} 
From the sine and cosine terms $x_2$ and $x_3$, 
we can compute the amplitude of annual seasonal changes in temperatures:
\[
    \sqrt{x_2^2 + x_3^2} = 22.02 \pm 0.15 ~\degF.
\]
In other words, on the hottest summer day we should expect the temperature to be
$22.02$ degrees warmer than the nominal value of $52.56$ degrees;  that is, $74.58$ degrees.  
Of course, this is a daily average---daytime highs can be expected to be
higher and nighttime lows lower by about the same amount.

Similarly, from the $x_4$ and $x_5$ sine and cosine terms, 
we can compute the amplitude of the temperature changes brought about by the solar-cycle:
\[
    \sqrt{x_4^2 + x_5^2} = 0.2887 \pm 0.156 ~\degF.
\]
The effect of the {\em solar cycle} is real but relatively small.

\subsection{Phase of the Sinusoidal Fluctuations}

Close inspection of the output shows that January 22 is nominally
the coldest day in the winter and July 24 is the hottest day of summer.
The plus/minus $2\sigma$-error on these estimates is less than half a day.
It is perhaps worth noting that the coldest days in the winter of 2011 turned
out to be January 23 and 24.

According to the LAD model,
February 12, 2007, was the day of the last minimum in the $10.7$ year solar cycle.
In this case, the plus/minus $2\sigma$-error on this estimate is about $400$
days.
It is well-known that the solar cycle had its last minimum in 2007 \cite{SW09}.
The correct extraction of the phase (and, in a later section, the period) 
of the solar cycle, which is
a small effect having an amplitude of only $0.29 ~\degF$, is strong
support for fidelity of the LAD model.


\begin{figure}
\begin{center}
\includegraphics[width=6.5in]{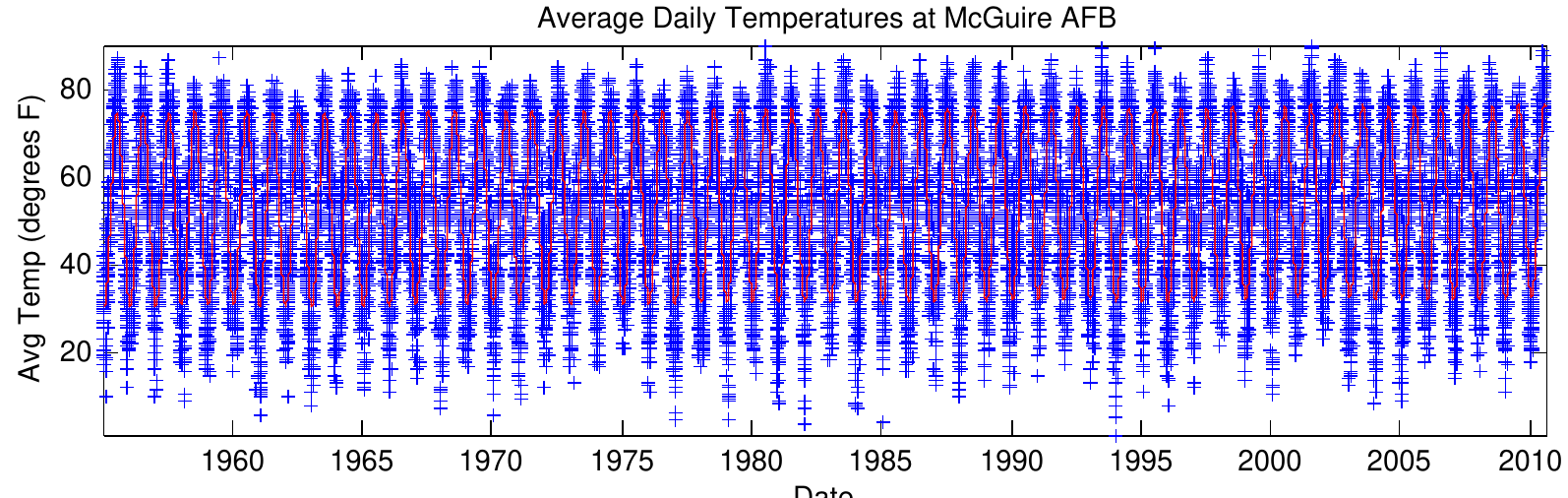} 
\end{center}
\caption{
Plot Showing Actual Data and Regression Curve. 
{\em Blue:} Average daily temperatures at McGuire AFB from 1955 to 2010. 
{\em Red:} Output from least absolute deviation regression model.
}
\label{fig3}
\vspace*{0.3in}
\end{figure}

\subsection{Visualizing the Results}

Figure \ref{fig3} shows a plot of all 20,309 data points.  Overlaid on these
data points is the solution of the LAD regression model.   
Daily and seasonal fluctuations completely dominate other effects.  It is impossible to
``see'' any linear warming trend or the solar cycle.

\begin{figure}
\begin{center}
\includegraphics[width=6.5in]{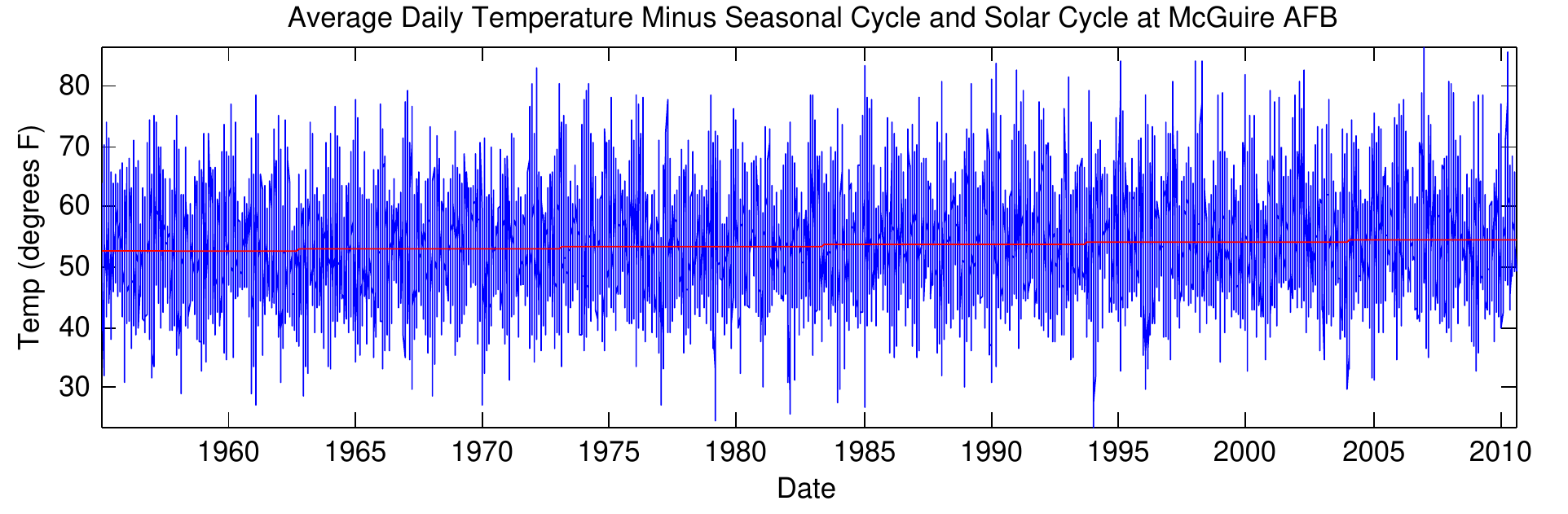} \\
\end{center}
\caption{
As before but with sinusoidal seasonal variation removed and
sinusoidal solar-cycle variation removed as well.
}
\label{fig4}
\vspace*{0.3in}
\end{figure}

Figure \ref{fig4} has the seasonal and solar-cycle variations removed.
Even this plot is noisy.  Of course, there are many days in a year and some
days are unseasonably warm while others are unseasonably cool.  It is not
uncommon for there to be an ``unseasonably warm'' day that is $20$ or even
sometimes $30$ degrees above seasonally adjusted averages.

\begin{figure}
\begin{center}
\includegraphics[width=6.5in]{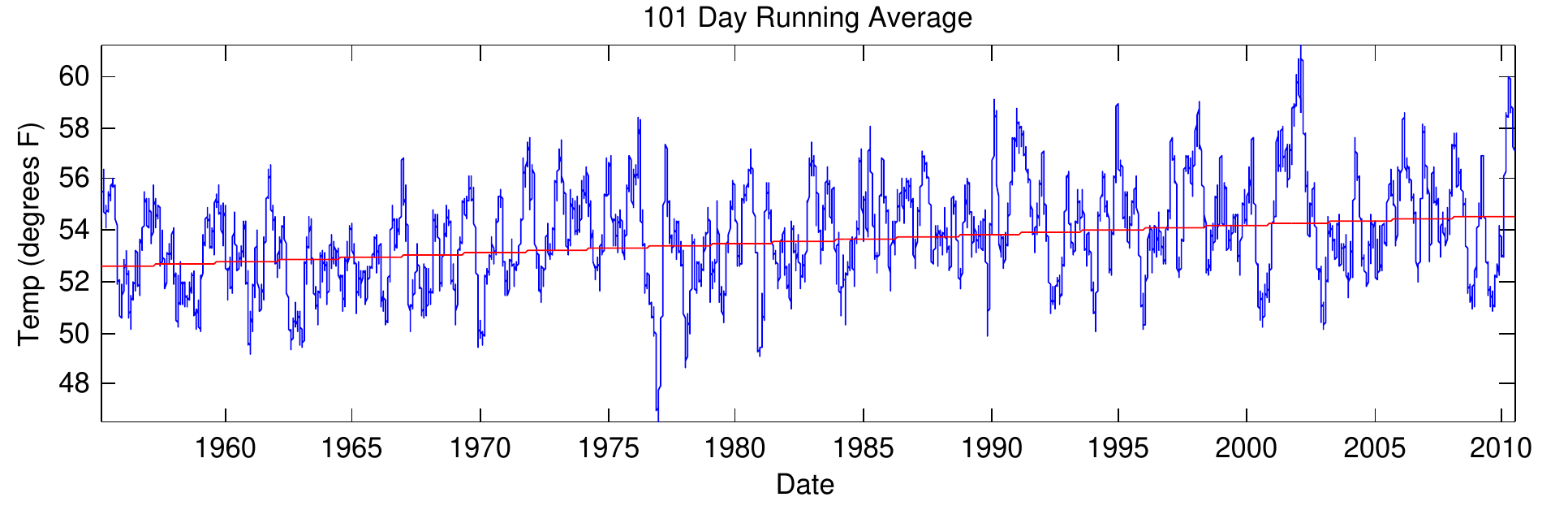} 
\end{center}
\caption{
Smoothed Seasonally Subtracted Plot. 
To smooth out high frequency fluctuations, we use $101$ day rolling averages of the data.
}
\label{fig5}
\vspace*{0.3in}
\end{figure}

Figure \ref{fig5} is derived from Figure \ref{fig4} by applying a $101$ day
rolling average to each data point.   The rolling average reduced the extreme
fluctuations to about $1/10$-th their original amplitude thus making the linear
warming trend quite apparent.  In NJ we have {\em local warming}.  

\section{Estimating the Period of the Solar Cycle} 

The length of the solar cycle is only approximately $10.7$ years \cite{WH91}.  We can modify
the model to predict, in addition to the parameters already being estimated,
the period of this cycle.  To do this, we introduce one new variable $x_6$ and
change the solar-cycle sine and cosine terms to read:
\[
	  x_4 \cos( 2 x_6 \pi d/(10.7 \times 365.25)) 
        + x_5 \sin( 2 x_6 \pi d/(10.7 \times 365.25)).
\]
If the unknown parameter $x_6$ is {\em fixed at $1$}, forcing the solar-cycle to have
a period of exactly $10.7$ years, then the problem reduces to the linear
programming problem considered earlier.  
If, on the other hand, we allow $x_6$ to vary, then the problem
is {\em nonlinear} and even {\em nonconvex} and therefore potentially harder to solve.  
However, nonlinear (local) optimization algorithms produce provably locally
optimal solutions ``near'' to the initially provided values for the variables.
Hence, for problems in which rough estimates of the optimal values are known,
one can expect such algorithms to produce the desired solution.  Such is the
case with the problem at hand.  The result is that the first six parameters
remain virtually unchanged and $x_6 = 0.992$ which translates to a $10.78$ year solar
cycle, in close agreement with the nominal value of $10.7$ years.
One might argue that it is unfair to initialize $x_6$ so that the starting point is
so close to the known correct solution.  Table \ref{tab3} shows the period of
the solar cycle obtained using various initial guesses.
\section{Least Squares Solution (Mean instead of Median)} \label{sec-lsr}
Suppose we change the objective to a sum of squares of deviations: \\[0.05in]
\begin{verbatim}
    minimize sumdev: sum {d in DATES} dev[d]^2;
\end{verbatim}
\vspace*{0.1in}

The resulting model is a {\em (nonlinear) least squares model}.  
The problem is no longer representable as a linear programming problem, but its
objective function is still convex and the problem is still easy
to solve using nonlinear optimization software.
Fixing $x_6$ to one (i.e., fixing the solar cycle to $10.7$ years), 
the problem becomes a linear least squares regression model that
can also be solved using any
number of statistical packages, such as R \cite{VS04}. 
The solution, however, is {\em sensitive} to outliers, which may or may not be
present.
Here's the output associated with the nonlinear least squares formulation:

\begin{eqnarray*}
x_0 &=& \phantom{-}52.6 ~\degF \\
x_1 &=& \phantom{-0}1.2 \times 10^{-4} ~^\circ \mbox{F/day}\\
x_2 &=&-20.3 ~\degF \\
x_3 &=& \;\;-7.97 ~\degF \\
x_4 &=& \;\;0.275 ~\degF \\
x_5 &=& \;\;0.454 ~\degF \\
x_6 &=& \;\;0.730 
\end{eqnarray*}

In this case, the rate of local warming is $4.37 ~\degF$ per century.
This number lies near the upper limit of the confidence interval given before. 
As Table \ref{tab3} shows, the model also produces a {\em wrong answer} for the period of the solar
cycle for each of the eleven values I used to initialize $x_6$ including $x_6 =
1$, which corresponds to a solar cycle of $10.7$ years.
It is well known that the period of the solar cycle
has been about $10.7$ years for the past few centuries.

\begin{table}
\begin{center}
\begin{tabular}{rrlrrlrrl}
\multicolumn{3}{c}{Initial period (in years)} & \multicolumn{6}{c}{Locally optimal period
	(in years)} \\[0.1in]
&&& \multicolumn{3}{c}{LAD} & \multicolumn{3}{c}{\qquad Least Squares} \\[0.04in]
\qquad\qquad& 6\phantom{.0}  &&& 6.64 &&\qquad\qquad& 14.65 \\
& 7\phantom{.0}  &&& 10.78  &&& 6.53 \\
& 8\phantom{.0}  &&& 10.78  &&& 6.53 \\
& 9\phantom{.0}  &&& 5.67  &&& 6.53 \\
& 10\phantom{.0} &&& 8.12  &&& 8.75 \\
& 10.7           &&& 10.78 &&& 14.65\\
& 11\phantom{.0} &&& -14.61  &&& 14.65 \\
& 12\phantom{.0} &&& 4.19  &&& 14.65 \\
& 13\phantom{.0} &&& -14.61  &&& 5.74 \\
& 14\phantom{.0} &&& 10.78  &&& 14.65 \\
& 15\phantom{.0} &&& -10.78 &&& $\infty$ \\[0.1in]
\end{tabular}
\end{center}
\caption{The derived period, in years, of the solar cycle computed using different initial
guesses for variable $x_6$ in the nonlinear versions of both the LAD and
least-squares optimization models.   In the table
values of $x_6$ are specified by giving the period of the sinusoid measured in years
(recall that $x_6 = 1$ corresponds to a $10.7$ year solar cycle).
Note that, for the least absolute deviations model, 
four of the eleven cases considered produced an answer close to the known answer
whereas none of them are very close using the least squares model.}
\label{tab3}
\end{table}
\section{Humidity}
It is well-known that higher temperatures imply more evaporation of water into
the atmosphere; that is, increased temperature implies increased humidity.
It turns out that the NOAA data sets contain not only temperature data but also
dew point data.  {\em Dew point} is the temperature below which water in the
atmosphere condenses out to form clouds/fog.  It is a quantitative measure of
the amount of water in the atmosphere---the more water vapor, the higher the
temperature must be for the air to hold that water
vapor.  It is common in summer months for there to 
be fog in the morning.  The temperature and the dew point match when it is
foggy.  One says that the {\em relative humidity} is $100\%$ at such times.  In 
the summer months at mid-latitudes, dew points are often up in the $60$ to $70$
degrees Fahrenheit range.  In the winter, obviously, the dew point is much
lower---winter air is significantly dryer than summer air.  Anyway, while weathermen
typically report humidity in terms of relative humidity, dew point
is a more direct measure of the amount of water vapor in the atmosphere.  

Given that NOAA tabulates this measure in the same data sets as used above, it
turns out to be almost trivial to modify the scripts to perform a dew point regression.  
What we discover is that, in 1955 at McGuire Airforce Base,
the nominal dew point (given by $x_0$) was $41.4 ~\degF$ and that on the
dampest summer days it was expected to be $22.6 ~\degF$ higher and 
on the driest winter days it was expected to be that much lower.
Furthermore, dewpoint is going up at a rate of $5.4 ~\degF$ per century---a rate
even greater than the rate that temperatures are rising, which means that
relative humidity is also increasing.

In the US, the National Weather Service reports a number called {\em heat index} on hot summer
days.   This index combines temperature and humidity into a single number on a
temperature scale.  However, this measure is fairly subjective as it is based on
relative humidity and aims to approximate how the temperature ``feels'' to a
person.  In Canada, a similar measure called {\em humidex} is used
(\cite{MR79,STR01}).  An
advantage of the humidex over heat index is that humidex is based on temperature
and dew point rather than temperature and relative humidity.   In fact, the
formula for humidex $H$ (in degrees Fahrenheit) is rather simple:
\[
    H = T + 6.11 e^{5417.7530(\frac{1}{273.16} - \frac{1}{D})} - 10, 
\]
where $T$ is temperature in degrees Fahrenheit and $D$ is dew point in degrees
Kelvin (of course, it is easy to convert degrees Fahrenheit to degrees Kelvin).

Again, the fact that NOAA tabulates dew point together with temperature
data makes it easy to produce values for the humidex.
What we discover is that, in 1955 at McGuire Airforce Base,
the nominal humidex (given by $x_0$) was $52.9 ~\degF$ and that on the
hottest/dampest summer days it was expected to be $30.8 ~\degF$ higher and 
on the coldest/driest winter days it was expected to be that much lower.
Furthermore, humidex is going up at a rate of $6.5 ~\degF$ per century---a rate
even greater than the rate that temperatures are rising.

\section{Going Global}
\begin{figure}
\begin{center}
\includegraphics[width=6.5in]{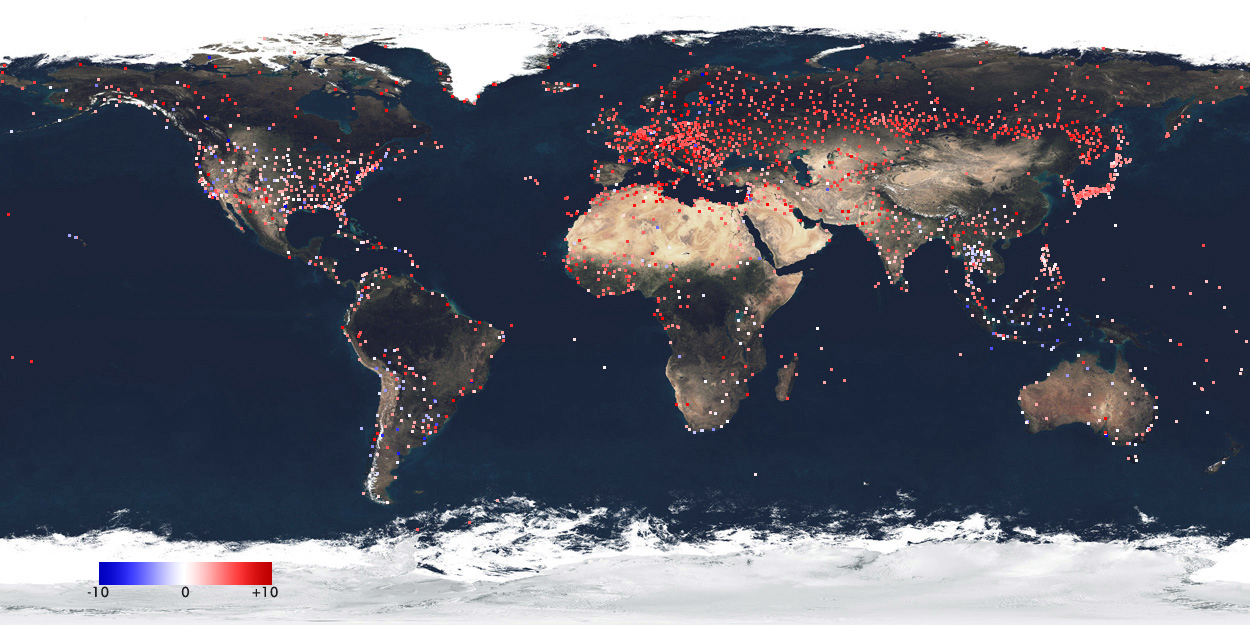}
\end{center}
\caption{
A map of $2399$ local warming results distributed around the globe.
In units of degrees Fahrenheit per century,
the mean of these values is $4.18$,
the median value is $4.53$, and
the standard deviation is $2.94$.
Dividing by the square root of the sample size, a confidence interval for the mean is
$ [ 4.12, 4.24 ] $.
}
\label{fig6}
\vspace*{0.3in}
\end{figure}

For fun, I applied the model to thousands of local weather stations
around the world to produce a global ``local warming map''.
Specifically, data files were downloaded from every NOAA weather station for which
collection commenced prior to Jan 1, 1955 and is currently in
ongoing.  There may be, and it turns out that there usually are, gaps in the
data---sometimes just a day here or there is missing but often there are multi-year
gaps.  Hence, the download script stipulated that the location must
have collected at least 3650 days of data (i.e., 10 years worth).
The resulting map is shown in Figure \ref{fig6}.

I should mention that no attempt was made to filter out bad data.
Also, seasonal variations are not sinusoidal in the tropics and so the simple
model presented earlier is not very good at tropical latitudes.

One can clearly see from Figure \ref{fig6} that warming is more pronounced in
the northern hemisphere.  Such an observation is consistence with the hypothesis
that warming over the last century is largely anthropogenic.

\section{Final Remarks} 

It is remarkable that the solar cycle can be seen and a warming trend can be
extracted from just one weather station's $55$-year dataset. 

In producing confidence intervals, one of our assumptions is wrong.   
In the bootstrap method, we assumed that the
temperature fluctuations from day to day are independent.  In reality they are
correlated over short time intervals---the correlation length is probably about
a week or so.  This error makes our confidence intervals too tight.  
If we know, or can guess, the correlation length $m$, then we can scale the
estimated standard deviation by the square root of $m$.
For example, if we assume that the correlation length is $9$ days, then the new
confidence interval for $x_1$ is 
$[1.38 ~\degF,\; 5.88 ~\degF]$ per century.

The original model can be improved in a few key ways.  First of all, the assumption
that the seasonal variation is sinusoidal is only an approximation---it
falls apart for data collection sites in the tropics where, in
principle, each year has two dates at which the Sun passes directly overhead and
two dates in between when the Sun is furthest (to the north/south) from passing overhead.  
Also, the linear
trend could be modeled as a function of global (or local) population density.
Over $55$ years, such a function is probably fairly, but certainly not exactly,
linear.   Hence, the basic models we have considered can be
improved in various ways.  Such improvements could inspire many 
interesting student projects.

\subsection{Getting the Data}
Since the NOAA data is archived in one year batches, 
I wrote a {\sc unix} shell script to grab the $55$
annual data files for McGuire and then assemble the relevant pieces of data into a
single file.  
Here is the shell script:

\vspace*{0.1in}
\noindent {\small http://www.princeton.edu/$\sim$rvdb/ampl/nlmodels/LocalWarming/McGuireAFB/data/getData.sh}
\vspace*{0.1in}

\noindent The resulting pair of data files that I used as input to my 
local climate model are posted at:

\vspace*{0.1in}
\noindent{\small http://www.princeton.edu/$\sim$rvdb/ampl/nlmodels/LocalWarming/McGuireAFB/data/McGuireAFB.dat}
\vspace*{0.1in}

and

\vspace*{0.1in}
\noindent{\small http://www.princeton.edu/$\sim$rvdb/ampl/nlmodels/LocalWarming/McGuireAFB/data/Dates.dat}
\vspace*{0.1in}

\bibliographystyle{siam}
\bibliography{../lib/refs}

{\bf Acknowledgments.} The author thanks 
Kurt Anstreicher,
Jianqing Fan,
J. Richard Gott,
Matthew Saltzman, and
Henry Wolkowicz
for helpful discussions about this work.

\end{document}